\documentstyle[twoside,fleqn,espcrc2,epsfig,psfig]{article}

\newcommand{\AmS}{{\protect\the\textfont2

  A\kern-.1667em\lower.5ex\hbox{M}\kern-.125emS}}

\title{Percolation and Critical Behaviour in SU(2) Gauge Theory
\thanks{The work has been supported by the TMR network ERBFMRX-CT-970122
and the DFG under grant Ka 1198/4-1.} 
}

\author{S. Fortunato, F. Karsch, P. Petreczky, H. Satz
\\\vskip 6pt
Fakult{\"a}t f{\"u}r Physik, 
    Universit{\"a}t Bielefeld,\\
    D-33501 Bielefeld, Germany} 
 
\begin{document}

\begin{abstract}
The paramagnetic-ferromagnetic transition
in the Ising model can be described as
percolation of suitably defined clusters.
We have tried to extend such picture to the confinement-deconfinement
transition of $SU(2)$ pure gauge theory,
which is in the same
universality class of the Ising model.
The cluster definition is derived 
by approximating $SU(2)$ by means of 
Ising-like effective theories. The geometrical transition
of such clusters turns out to describe successfully
the thermal counterpart
for two different lattice regularizations of $(3+1)$-$d$ $SU(2)$.
\end{abstract}

\maketitle

\section{INTRODUCTION}
The search 
for a geometrical description 
of phase transitions has stimulated
a lot of interest over the last decades. 
Such descriptions are 
simple and elegant and allow to
justify the relationship between
critical phenomena and geometry which 
seems to play a key role in
important aspects of 
critical behaviour, like the universality of 
the critical indices.

In particular, percolation theory \cite{stauff,grimm} 
represents an ideal framework 
for the required formulation. 
Percolation takes place
when geometrical clusters, formed by 
elementary objects of some system, give rise to
an infinite network
(percolation cluster) which spans the whole system.
The percolation phenomenon
turns out to have astonishing analogies with
ordinary second order thermal phase transitions:
\vskip2mm
{$\bullet$} the behaviour of the
percolation variables at criticality is also
described by simple power laws, with corresponding exponents;
\vskip2mm
{$\bullet$} the values of the exponents, related to each other by simple scaling relations,
are fixed only by the number of space dimensions of the system
at study, regardless of its structure and of the special type
of percolation process one considers.
\vskip2mm
The spontaneous symmetry breaking of the Ising model
can be indeed described by means of percolation
of site-bond clusters \cite{Co80}. Since 
the critical behaviour of 
$SU(2)$ pure gauge theory is in the same universality class as the
Ising model \cite{svet,eng}, 
an equivalent percolation picture for 
the confinement-deconfinement transition of $SU(2)$ 
has been recently proposed \cite{santo2D}.
Unfortunately, the approach suggested in \cite{santo2D}
is valid only in the strong coupling limit 
and the problem of finding a general description remains open.

In this work, we will show that it is possible to  
define the clusters we need by constructing 
suitable {\it effective theories} for $SU(2)$ which
admit an equivalent percolation picture. The effective models
are Ising-like spin models with only spin-spin interactions,
and we will show that they provide a working 
percolation description of the critical behaviour
of $(3+1)$-$d$ $SU(2)$ both in
the strong coupling limit ($N_{\tau}=2$) 
and for $N_{\tau}=4$, a case which
approaches 
the weak coupling region.

\section{PERCOLATION AND CRITICAL BEHAVIOUR IN THE ISING MODEL}

The first percolation studies 
on the Ising model concerned  
the ordinary magnetic domains, i.e. 
clusters formed by nearest-neighbour
spins of the same sign.
In two dimensions
such clusters happen indeed to percolate at the thermal
critical point $T_c$ \cite{connap}. Nevertheless, the values
of the critical exponents differ from the 
corresponding Ising values \cite{sykes}. 
In three dimensions, 
the magnetic domains of 
the spins oriented in the direction of the magnetization percolate at any temperature;
the domains formed by the spins opposite to the magnetization
percolate for $T{\geq}T_p$, 
with $T_p{\neq}T_c$ \cite{krumb}.

It was then clear that the magnetic domains were 
too big to reproduce the thermal behaviour of the Ising model.
The fact that two nearest neighbour
spins of the same sign are always connected to each other 
implies that the clusters fail to reproduce
the correct spin correlation because of geometrical effects.
For this reason,
Coniglio and Klein \cite{Co80} suggested that
two like-signed nearest neighbouring spins
belong to the same cluster with a certain {\it bond probability}
\begin{eqnarray}
p=1-\exp(-2J/kT) 
\label{io}
\end{eqnarray}
($J$ is the Ising coupling, $T$ the temperature).
The clusters introduced by 
Coniglio and Klein are thus
{\it site-bond} clusters:
their percolation transition is indeed equivalent
to the magnetization transition of the Ising model.
 
As a matter of fact, this result is 
valid for a wide class of 
models. For example, if 
a theory is characterized just by ferromagnetic 
spin-spin interactions, the percolation picture
of Coniglio and Klein can be trivially
extended by introducing a bond 
between each pair of interacting spins,
and a relative bond probability
\begin{eqnarray}
\label{tu}
p_i=1-\exp(-2J_i/kT), 
\end{eqnarray}
where $J_i$ is the coupling 
associated to the $i$-interaction of the theory.

This generalization is at the basis of 
our work.

\section{POLYAKOV LOOP PERCOLATION IN SU(2) GAUGE THEORY}

Let us now consider the case of $SU(2)$ gauge theory.
We have again a $Z(2)$ symmetric variable,
the {\it Polyakov loop} $L$, whose 
thermal lattice average is the order parameter of
the confinement-deconfinement transition. 
Therefore, the $SU(2)$ configurations are similar to
the Ising ones, with the Polyakov loop
playing the role of the spin variable.
Also here we will have 
regions in which $L$ has the same sign,
analogous to the magnetic domains of the Ising model.
However, to identify the physical clusters, 
as we have seen above, we need to 
introduce some bond probability. 
This is a difficult problem, since the 
$SU(2)$ lattice action cannot be expressed in terms of the 
Polyakov loop $L$.
In order to derive the correct bond weights, in fact, 
it is necessary to know how the "gauge spins", i.e.
the Polyakov loops, interact with each other.

One way to face the problem is 
to approximate $SU(2)$ by means of effective theories
which are easy to handle.
In \cite{santo2D} one derived an effective theory
from a strong coupling expansion of $SU(2)$.
That method thus cannot be applied to the weak coupling case.
 
To find a general procedure, we shall construct the effective theory
starting not from the $SU(2)$ Lagrangian, but from
the Polyakov loop configurations.
If it were possible to find an effective model 
which reproduces well the Polyakov loop configurations
and has an equivalent percolation formulation, the problem
would be solved.
The Polyakov loop, unlike the Ising spins, is a continuous variable, 
but we shall consider the Ising-projected configurations,
i.e. the distributions of the {\it signs} of the Polyakov loops.
This is done assuming that the $Z(2)$ symmetry is the 
only relevant feature at the basis of the critical behaviour.
Our ansatz for the effective theory will be an Ising-like model 
with just spin-spin interactions.
Its Hamiltonian  
${\cal H}(s)$ is 
\begin{eqnarray}
\label{ansatz}
{\cal H}(s)\,=\,-J_{1}\sum_{NN}s_is_j\,-\,J_{2}\sum_{NNN}s_ks_l\,
-\,etc.\,,
\end{eqnarray}
where the distance between coupled spins increases
progressively starting from the simple nearest-neighbour ($NN$) case
($NNN=$next-to-nearest, and so on).
We have seen in the previous
section that, if $J_i>0\,\, \forall i$, such a model has a simple percolation picture a la 
Coniglio-Klein. 

The relationship between $SU(2)$ and
the effective model 
is established through the equation 
\begin{eqnarray}
e^{-{\cal H}(s)/kT}\,=\,\int\,[dU]\prod\limits_{\bf n}\delta[s_{\bf n}, sgn(L_{\bf n})]
\,e^{S_{SU2}},
\end{eqnarray}
where $L_{\bf n}$ is the 
value of the Polyakov loop at the spatial point {\bf n} and 
$S_{SU2}$ the $SU(2)$ lattice action. 

The couplings of the effective theory are calculated 
by solving a set of Schwinger-Dyson equations, a method which
gives good results
in Monte Carlo renormalization group studies
of field theories \cite{okawa,oka2}.
If the couplings
turn out to be all positive, we can use them to 
calculate the corresponding bond weights we need
to build the clusters
in the original Polyakov loop configurations. Because of the 
expression (\ref{tu}) for the bond weights, with this approach
only the configurations of the signs of the Polyakov loops
are relevant for the cluster building.

\section{RESULTS FOR $(3+1)$-$d$ SU(2)}

We performed simulations of $(3+1)$-$d$
$SU(2)$, focusing on the cases
$N_{\tau}=2$ and $N_{\tau}=4$
($N_{\tau}$ is the number of lattice spacings in the 
temperature direction).
We found that the two cases can be 
indeed approximated by models like (\ref{ansatz}),
with 15 and 19 ferromagnetic couplings, respectively.
The distances between the spins in the longest-range
interactions, in lattice units,
are $\sqrt{17}$ for $N_{\tau}=2$ and $\sqrt{27}$ 
for $N_{\tau}=4$. The relative strength of such interactions
compared to the nearest-neighbour couplings is about $1/1000$
in both cases, which shows how rapidly the couplings decrease with the distance
between the spins.

Both for $N_{\tau}=2$ and 
for $N_{\tau}=4$ we used four different lattice sizes, 
in order to be able to extract 
the results by applying finite size scaling 
techniques.

\begin{figure}[htb]
\epsfig{file=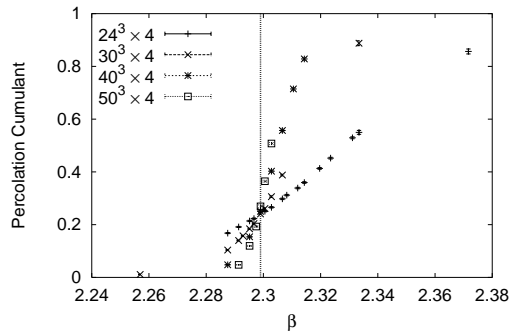,width=75mm}
\vspace{-12mm}
\caption{Percolation cumulant as a function of $\beta$ for 
four lattice sizes: $N_{\tau}=4$.}
\label{due}
\end{figure}

\begin{figure}[htb]
\epsfig{file=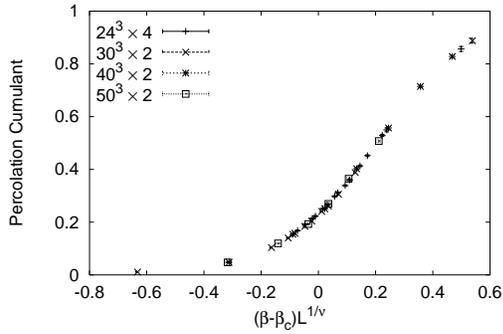,width=75mm}
\vspace{-12mm}
\caption{Rescaling of the percolation cumulant 
curves of Fig.\ref{due}, using the 3D Ising exponent $\nu=0.630$.}
\label{qua}
\end{figure}

To locate the critical point of the
percolation transition, we made use of the 
{\it percolation cumulant}. Such variable is determined
simply by counting, for a fixed lattice size and 
a value of the $SU(2)$ coupling $\beta=4/g^2$, how many 
configurations contain a spanning cluster.
This number, divided by the total number of configurations,
is directly a scaling function,
analogous to the Binder cumulant 
in continuous thermal phase transitions \cite{Bin}.
\begin{figure}[htb]
\epsfig{file=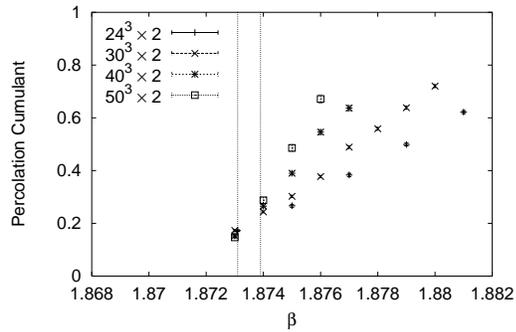,width=75mm}
\vspace{-12mm}
\caption{Percolation cumulant as a function of $\beta$ for 
four lattice sizes: $N_{\tau}=2$.}
\label{uno}
\end{figure}
Figs. \ref{uno} and \ref{due} show the 
behaviour of the percolation cumulant
curves as a function of $\beta$, for $N_{\tau}=2$
and $4$, respectively. The curves cross remarkably at
the same point, within errors,
in good agreement with the thermal thresholds, indicated
within one standard deviation by the dashed lines. 
The percolation cumulant gives the critical point $\beta_{c}$
but also the critical exponent $\nu$.
In fact, if we rewrite the percolation cumulant as a function
of $(\beta-\beta_{c})L^{1/\nu}$ ($L$ is the linear spatial dimension
of the lattice), we should get the same curve
for each lattice size. 
Fig. \ref{tre} and \ref{qua} 
show the rescaled percolation cumulant using 
for $\beta_{c}$ the position of the crossing points
and for the exponent $\nu$ the  
3D Ising value. 
In both cases, the curves coincide.
\begin{figure}[htb]
\epsfig{file=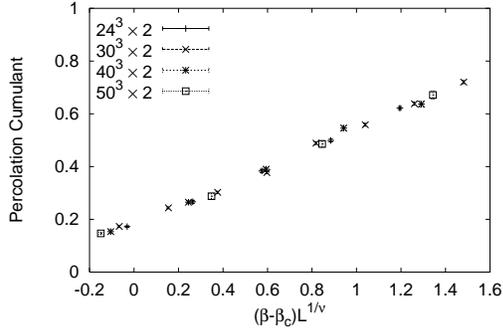,width=75mm}
\vspace{-12mm}
\caption{Rescaling of the percolation cumulant 
curves of Fig.\ref{uno}, using the 3D Ising exponent $\nu=0.630$.}
\label{tre}
\end{figure}
\vskip-0.05cm
To complete the investigation of
the geometrical transition, we determined the ratio
of critical exponents 
$\gamma/\nu$ from a standard finite size scaling analysis.
We were not able to evaluate 
the ratio $\beta/\nu$, because of large fluctuations
of the percolation order parameter, the {\it percolation strength} $P$,
around the critical coupling. 
The final results are listed in Tables \ref{unta} and \ref{duta}, where we give, 
for comparison, also the critical indices of
the thermal transition. The agreement is good in both cases.
\begin{table}[hbt]
\newlength{\digitwidth} \settowidth{\digitwidth}{\rm 0}
\catcode`?=\active \def?{\kern\digitwidth}
\caption{Percolation critical indices for 
$(3+1)$-$d$ $SU(2)$, $N_{\tau}=2$. The thermal indices are taken from
\cite{fortuna}.}
\label{unta}
\begin{tabular*}{73.8mm}{|c|c|c|c|}
\hline
& Crit. Point& $\gamma/\nu$& $\nu$\\
\hline
Perc.& $1.8734(2)$
        &$1.977(15)$ & $0.628(10)$  \\\hline 
Therm. &$1.8735(4)$ &$1.953(14)$  &$0.630(9)$ \\\hline
        \end{tabular*}
\end{table}
\vskip-1cm
\begin{table}[hbt]
\catcode`?=\active \def?{\kern\digitwidth}
\caption{Percolation critical indices for 
$(3+1)$-$d$ $SU(2)$, $N_{\tau}=4$. The thermal indices
are taken from \cite{eng}.}
\label{duta}
\begin{tabular*}{74.5mm}{|c|c|c|c|}
\hline
& Crit. Point& $\gamma/\nu$& $\nu$\\
\hline
Perc.& $2.2991(2)$
        &$1.979(15)$ & $0.629(8)$  \\\hline 
Therm. &$2.29895(10)$ & $1.944(13)$ & $0.630(11)$ \\\hline
        \end{tabular*}
\end{table}
\newpage
\section{CONCLUSIONS}

We have shown that the deconfinement transition of SU(2) 
gauge theory in 3+1 dimensions can be described in terms of
Polyakov Loop percolation. The procedure we used is, 
necessarily, approximate. However, in comparison with
the method illustrated in 
\cite{santo2D}, it has the advantage
to be applicable also to the weak coupling limit
of $SU(2)$.


\vskip2cm

\begin{thebibliography}{9}

\bibitem{stauff} D. Stauffer, A. Aharony,
  {\it Introduction to Percolation Theory},
      Taylor {\&} Francis, London 1994.

\bibitem{grimm} G. Grimmett, {\it Percolation}, Springer-Verlag, 1999.

\bibitem{Co80} A. Coniglio, W. Klein, J.Phys. A: Math. Gen. {\bf 13}, 
  2775-2780 (1980). 

\bibitem{svet} B. Svetitsky, L. Yaffe, Nucl. Phys. B {\bf 210} [FS6],
  423 (1982); 
  L. Yaffe,
      B. Svetitsky, Phys. Rev. D {\bf 26}, 963 (1982). 
      B. Svetitsky, Phys. Rep. {\bf 132},
      1 (1986).

\bibitem{eng} J.\ Engels et al., Phys. Lett. B {\bf 365}, 219 (1996).

\bibitem{santo2D} S. Fortunato, H. Satz, Phys. Lett. B {\bf 475}, 311 (2000).

\bibitem{connap} A. Coniglio, C. R. Nappi, F. Peruggi, L. Russo, Commun. Math. Phys. {\bf 51}, 315 (1976).

\bibitem{sykes} M. F. Sykes, D. S. Gaunt, J. Phys. A {\bf 9}, 2131 (1976).

\bibitem{krumb} H. M\"uller-Krumbhaar, Phys. Lett. A {\bf 48}, 459 (1974).

\bibitem{okawa} M. Okawa, Phys. Rev. Lett. {\bf 60}, 1805 (1988).

\bibitem{oka2} A. Gonzalez-Arroyo, M. Okawa, Phys. Rev. Lett. {\bf 58}, 2165 (1987).

\bibitem{Bin} K. Binder, D. W. Heermann, {\it Monte Carlo simulations in
Statistical Physics, An Introduction}, Springer-Verlag 1988. 

\bibitem{fortuna} S. Fortunato, H. Satz, Proceedings of the $3^{rd}$ Catania Relativistic
  Ion Studies (CRIS 2000),
Acicastello (Italy).

\end{thebibliography}
\end{document}